\documentstyle[manuscript,prl,aps]{revtex}

\begin{document}
\draft
\title{Ultrafast optical nonlinearity in quasi-one-dimensional 
       Mott-insulator ${\rm Sr_2CuO_3}$ }
\author{T. Ogasawara$^{1}$, M. Ashida$^{2}$, N. Motoyama$^{1}$, 
        H. Eisaki$^{1}$, S. Uchida$^{1}$, Y. Tokura$^{1}$, 
        H. Ghosh$^{3}$, A. Shukla$^{2,3}$, S. Mazumdar$^{3}$, 
        M. Kuwata-Gonokami$^{1,2}$\cite{auth}}
\address{$^{1}$Department of Applied Physics, the University of Tokyo,\\
         7-3-1, Hongo, Bunkyo-ku, Tokyo, 113-8656, Japan}
\address{$^{2}$Cooperative Excitation Project, Japan Science and
         Technology Corporation (JST),\\
         3-2-1 Sakato, Takatsu-ku, Kanagawa 213-0012, Japan}
\address{$^{3}$Department of Physics and The Optical Sciences Center,\\
         University of Arizona, Tuscon, AZ85721, USA}
\date{\today}
\maketitle

\begin{abstract}
We report strong instantaneous photoinduced absorption in the
quasi-one-dimensional Mott insulator ${\rm Sr_{2}CuO_{3}}$ in the IR
spectral region. The observed photoinduced absorption is to an even-parity
two-photon state that occurs immediately above the absorption edge.
Theoretical calculation based on a two-band extended Hubbard model explains
the experimental features and indicates that the strong two-photon
absorption is due to a very large dipole-coupling between nearly degenerate
one- and two-photon states. Room temperature picosecond recovery of the
optical transparency suggests the strong potential of ${\rm Sr_{2}CuO_{3}}$
for all-optical switching.
\end{abstract}


Nonlinear optical materials with large nonlinear coefficient, fast
response-time, low loss and operability at room-temperature will be
indispensable for next generation high speed network systems in which clock
recovery and buffering at terabit/second rate will be performed by
all-optical switches \cite{stegman}. These devices should operate in the
transparent region below the fundamental band edge, where optical
nonlinearity is mainly associated with two-photon absorption (TPA). Real
carrier excitation, inevitable due to TPA, is considered as the major
limiting factor here. Although terabit/second all-optical switches have been
already demonstrated with conventional inorganic band semiconductors\cite
{nakamura}, considerable effort is necessary to overcome the difficulty
associated with long carrier lifetime. Large optical nonlinearity and
sub-picosecond response times are observed in organic $\pi $-conjugated
polymers \cite{chemla,messier}, but further improvements in sample quality
and morphology would be required for actual applications. Parallel
development of novel nonlinear optical inorganic materials that on the one
hand possess the large nonlinearity and ultrafast recovery times of the
organics, and the intrinsic robustness and superior thermal conductivity of
inorganics on the other, is highly desirable. To date, no such inorganic
semiconductor is known. We report in the present Letter the observation of
large ultrafast optical nonlinearity in a novel strongly correlated
inorganic semiconductor that is intrinsically different from conventional
inorganic band semiconductors. The specific material we have studied is $
{\rm Sr_{2}CuO_{3}}$.

${\rm Sr_{2}CuO_{3}}$ is a quasi-1D Cu-O linear chain compound, whose
structure is shown in Fig. \ref{evo}(a). The material is a prototypical
strongly correlated charge-transfer insulator within the
Zaanen-Sawatzky-Allen scheme \cite{ZSA}. Optical absorption in ${\rm 
Sr_{2}CuO_{3}}$ involves charge-transfer (CT) excitation of a Cu-hole to an
O-site, with an absorption band-edge at $\sim $ 1.6 eV and the band maximum
at 2 eV (Fig. \ref{evo}(b)). In addition to the high energy charge
excitations, there exist low energy spin excitations in ${\rm Sr_{2}CuO_{3}}$
, which correspond to those of a nearly ideal 1D Heisenberg chain with
intrachain exchange integral $J\sim 2000-3000\;{\rm K}$\cite
{suzuura,motoyama}. The interchain coupling in this system is rather weak,
as is evidenced by the occurrence of 3D long-range antiferromagnetic
coupling only below $T_{N}=5\;{\rm K}$ \cite{keren}.

We present here the results of time-resolved femtosecond pump-probe
measurements on a single crystal of ${\rm Sr_2CuO_3}$, grown by the
traveling-solvent floating-zone method \cite{motoyama}. A thin flake with a
thickness of $L \sim 50\;{\rm \mu m}$ was cleaved out with the $bc$-plane
for transmission measurement. Optical pulses are provided by a system based
on amplified mode-locked Ti:sapphire laser and optical parametric generator
supplemented with sum and difference frequency generation. This system
generates pulses with the photon energy centered at 0.2 -- 2.0 eV and with
temporal width $\sim 200\;{\rm fs}$. In our experiment, we measure the
differential transmission $\Delta T/T$ (where $T$ is the transmission in the
absence of the pump beam), as a function of the delay time of the probe with
respect to the pump.

Fig. \ref{evo}(c) shows the temporal evolution of $\Delta T/T$ measured at $
10\;K$ for different pump ($\omega _{{\rm pump}}$) and probe ($\omega _{{\rm 
probe}}$) photon energies. The intensity of the pump pulse is $\sim 2\;{\rm 
GW/cm}^{2}$. The intensity of the probe pulse is one to two orders of
magnitude smaller. One can observe from Fig. \ref{evo}(c) that the
differential transmission consists of two components. The first component
manifests itself as a sharp peak-like structure with temporal width of about 
$200{\rm \;fs}$, which is given by the overlapping of pump and probe pulses.
This component, which is pronounced when the pump photon energy is tuned
below the absorption edge of $\sim 1.6\;{\rm eV}$, is due to coherent
interaction between the pump and probe pulses. At $\omega _{{\rm pump}}$= $
3.1\;{\rm eV}$, which is far above the absorption edge, this coherent
interaction is suppressed by the strong linear absorption, and
correspondingly, the peak-like structure in the pump-induced transmission
vanishes. At $\omega _{{\rm pump}}$ = $1.55\;{\rm eV}$, the strong group
velocity dispersion near the absorption edge gives rise to the increase in
the temporal width of the peak-like structure up to approximately $1\;{\rm ps
}$.

The second component of the transmission change manifests itself as an
exponentially decaying tail with characteristic time of about $2\;{\rm ps}$.
One can observe from Fig. \ref{evo}(c) that magnitude of this component
decreases with decreasing of the pump photon energy and, correspondingly,
with the pump absorption coefficient (see Fig. \ref{evo}(b)). These
experimental findings indicate that this relatively long-lived component of
the transmission change can be associated with real excitation of the CT
exciton, whose decay time is of similar magnitude.

These features in the temporal behavior of the pump-induced transmission
remain at room temperature. We find that the magnitude of the effect is
proportional to the pump intensity up to $\sim 5\;{\rm GW/cm}^{2}$,
indicating the pump-induced transmission change to be a third order optical
process.

We describe $\Delta T/T$ in terms of the pump-induced change in the
absorption coefficient $\Delta \alpha =-\ln (1+\Delta T/T)/L$, where $L$ is
the crystal length, which consists of the instantaneous (peak) and
exponentially decaying (tail) components, respectively: $\Delta \alpha
=\Delta \alpha _{{\rm peak}}+\Delta \alpha _{{\rm tail}}$. Fig. \ref
{probe_dep}(a) shows the spectrum of TPA coefficient $\beta =\Delta \alpha _{
{\rm peak}}/I_{{\rm pump}}$ as a function of $\omega _{{\rm probe}}$ for
pump photon energies of $0.7\;{\rm eV}$, $1.1\;{\rm eV}$ and $1.55\;{\rm eV}$
. One can observe from Fig. \ref{probe_dep}(a) that the smaller the $\omega
_{{\rm pump}}$, the larger is the probe photon energy $\omega _{{\rm probe}}$
at which the maximum of the pump-induced coherent absorption is obtained.
Moreover, it is seen that $\beta $ is maximum at $\omega _{{\rm probe}
}\approx 2.1\;{\rm eV}$ -- $\omega _{{\rm pump}}$ at all pump photon
energies. This strongly suggests that the peak structure in Fig. \ref{evo}
(c) is due to a two-photon allowed (one-photon forbidden) state at $\sim $
2.1 {\rm eV}. Alternative mechanisms, which may be responsible for the
observed instantaneous photoinduced absorption change, can in principle
involve fifth- and higher order effects associated with real
carriers excitation due to two-photon absorption of the pump or with the
electron-hole plasma created by the intense pump pulse \cite{comment on EMT}
. However, the observed linear dependence of $\Delta \alpha _{{\rm peak}}$
on the pump intensity suggests that the instantaneous photoinduced
absorption is essentially of the third-order in the light field. This
observation along with the strong dependence of the pump-induced
transmission spectra on $\omega _{{\rm pump}}$ precludes mechanisms of
optical nonlinearity other than TPA.

To clarify this further we plot  $\beta $ in Fig. \ref{probe_dep}(b) as a
function of $\omega _{{\rm pump}}+\omega _{{\rm probe}}$. One can readily
observe that maximum of the pump-induced coherent absorption takes place at $
\omega _{{\rm pump}}+\omega _{{\rm probe}}=2.1{\rm \;eV}$ for all pump
photon energies. The dashed line on Fig. \ref{probe_dep}(b) represents the
linear absorption spectrum indicating that one- and two-photon allowed bands
nearly overlap.

In order to clarify the origin of the observed TPA band in Fig. \ref
{probe_dep}(b) we consider the Cu-O chain within the two-band extended
Hubbard model \cite{imada}, 
\begin{equation}
H=\sum_{i}U_{i}n_{i\uparrow }n_{i\downarrow
}+V\sum_{i}n_{i}n_{i+1}+\sum_{i,\sigma }(-1)^{i}\epsilon
n_{i}-t\sum_{i,\sigma }(c_{i\sigma }^{\dagger }c_{i+1\sigma }+c_{i+1\sigma
}^{\dagger }c_{i\sigma })  \label{eq_1}
\end{equation}
where $c_{i\sigma }^{\dagger }$ creates a hole with spin $\sigma $ on site $i
$, $n_{i\sigma }=c_{i\sigma }^{\dagger }c_{i\sigma }$, and $
n_{i}=\sum_{\sigma }n_{i\sigma }$. The parameter $t$ is the transfer
integral between $Cu$ and $O$ sites, $2\epsilon =\epsilon _{O}-\epsilon _{Cu}
$, where $\epsilon _{O}$ and $\epsilon _{Cu}$ are the site energies of the $O
$ and $Cu$ sites, $U_{i}$ is the on-site Coulomb repulsion between two holes
on $Cu$ and $O$ sites ($U_{Cu}\neq U_{O}$) and $V$ is the Coulomb repulsion
between holes occupying neighboring $Cu$ and $O$ sites. For large $U_{i}$
and realistic $\epsilon >0$, the holes occupy predominantly the $Cu$ sites
in the ground state, which can thus be represented by the hole occupancy
scheme ..101010...., where 1(0) represents an occupied(unoccupied) site.
From the nature of the current operator $\hat{\jmath}=-it\sum_{i,\sigma
}(c_{i\sigma }^{\dagger }c_{{i+1}\sigma }-c_{{i+1}\sigma }^{\dagger
}c_{i\sigma })$, the one-photon optical state has the form (...10100110...)
-- (...10110010...), i.e., the odd parity linear combination of the two
configurations that are reached by one application of the current operator
on the ground state. The even parity ``plus'' linear combination of the same
two configurations is a two-photon state, which for strong correlations is
nearly degenerate with the one-photon state. This near degeneracy is
expected to lead to very large transition dipole coupling between the one-
and two-photon states.

To confirm the above conjecture we have performed exact numerical
calculations for finite periodic rings of 12 sites (6 $Cu$ and 6 $O$) within
Eq.\ref{eq_1}, for parameters $|t|$ = 1 \-- 1.4 eV, $U_{Cu}$ = 8 \--10 eV, $
U_O$ =4 \-- 6 eV, $V$ = 0 - 2 eV, and $\epsilon$ = 1 -- 2 eV. Because of the
large Hubbard $U$ the relevant wavefunctions are strongly localized, and
even the 12-site periodic ring can give semi-quantitative results. In all
cases, the transition dipole coupling between the one- and the two-photon
states is one to two orders of magnitude larger than that between the ground
state and the one-photon state. The third order nonlinear optical
coefficient, $\chi^{(3)}(-\omega;\omega,\omega,-\omega)$, whose imaginary
component gives the TPA, is now calculated from the energies and transition
dipole couplings\cite{Boyd}. In Fig. \ref{calc} we have shown the calculated
absorption and TPA spectra in arbitrary units, for one set of parameters for
our finite system. The inset shows the energies of the one- and two-photon
states and the transition dipole couplings that are used in the calculation.

At the maximum of the TPA band, the magnitude of the experimental TPA
coefficient $\beta $ is $\sim 150\;{\rm cm/GW}$ (Fig. \ref{probe_dep}(b)),
which corresponds to ${\rm Im}\chi ^{(3)}\sim 10^{-9}$ {\rm esu}. This value
is larger than that predicted from the gap-dependent scaling law derived for
conventional semiconductors\cite{stryland} by one order of magnitude, and is
comparable to that of conventional 1D-structured materials. The scaling law 
\cite{stryland} is inapplicable to Mott-insulators, in which the origin of
the optical nonlinearity is the very large dipole coupling between nearly
degenerate excited states of opposite parities. The mechanism of optical
nonlinearity here is related to that in the $\pi $-conjugated polymers,
which are described within the {\it one}-band extended Hubbard model and in
which also there occurs very large dipole coupling between the optical state
and a two-photon state slightly higher in energy \cite{chandross}. Not
surprisingly, the magnitude of $\chi ^{(3)}$ in ${\rm Sr_{2}CuO_{3}}$ is
therefore comparable to some of the best organic materials \cite
{lawrence,bubeck}. The intensity dependent refractive index $n_{2}$,
obtained by a Kramers-Kronig transformation of the TPA data, is $\sim $ 10$
^{-7}$ -- 10$^{-6}$ cm$^{2}$/MW, also comparable to the organics \cite
{lawrence}.

In addition to the large magnitude, the excitonic nonlinearity in ${\rm Sr}
_{2}{\rm CuO}_{3}${\rm \ }is featured by picosecond response, whose
characteristic time is given by the decay constant of the tail-like
component in Fig. \ref{evo}(c). Such an ultrashort response indicates the
existence of a fast non-radiative relaxation channel. Although detailed
discussion of relaxation mechanisms of the optical excitation in ${\rm Sr}
_{2}{\rm CuO}_{3}$ is beyond the scope of this paper, we suggest that the
ultrafast ground state recovery is related to the occurrence of spin
excitations below the optical gap\cite{matsuda1,matsuda2}. The low energy
excitations of these system are the gapless spin excitations (spinons) of
the uniform one-dimensional antiferromagnetic Heisenberg chain. These spin
excitations are optically silent, and have an overall bandwidth of $\sim $ 1
eV \cite{suzuura,motoyama}. It is then conceivable that following the
relaxation of the CT exciton to the highest energy spin excited states there
occur further fast intra-spinon-band relaxation through the emission of
multiple phonons and spinons. Because of the absence of such midgap states
in the conventional semiconductors similar non-radiative processes would be
absent there.

The room temperature ultrafast ground state recovery implies a high
potential of ${\rm Sr_{2}CuO_{3}}$ in the ultrafast optoelectronics and,
specifically, in all-optical switching devices. In order to estimate this
potential in terms of the maximum available repetition rate, we have
performed a double-pulse experiment at room-temperature. Fig. \ref{double}
shows the $\Delta T/T$ induced by two pump pulses with wavelength of $1400\;
{\rm nm}$. The transmission change was measured at $1200\;{\rm nm}$, which
is around the optical fiber communication wavelength. The second pump pulse,
applied 2 ps after the first, induces nearly the same transmission change as
the first. However, comparing with the single-pulse response (not shown),
the transmission change in tail part is accumulated from pulse to pulse,
leading to the limitation on the repetition rate. If $\Delta \alpha _{{\rm 
tail}}$ and $\Delta \alpha _{{\rm peak}}$ are the tail and peak components
of photoinduced absorption and $f$ is the pulse repetition rate, the tail
component of the induced absorption in the steady-state regime is $\delta
\alpha _{{\rm tail}} =\Delta \alpha _{{\rm tail}}/[1-\exp (-1/\tau f)]$.
Therefore, the maximum available repetition rate can be estimated from a
natural criterion of the operability of the system, $\delta \alpha _{{\rm 
tail}}=\Delta \alpha _{{\rm peak}}$, which gives $f_{{\rm max}}=(\Delta
\alpha _{{\rm peak}}/\Delta \alpha _{{\rm tail}})\tau ^{-1}$. >From Fig. \ref
{double}, we obtain $f_{{\rm max}}\approx 10^{13} {\mbox s}^{-1}$ for our
sample, i.e., it can be used as a nonlinear optical medium with operability
of several terabits per second.

In summary, we find that ${\rm Sr_{2}CuO_{3}}$ exhibits strong
nonlinearity and picosecond recovery of optical trasparency.
Theoretical calculations indicates that the nonlinearity of the quasi-1D
cuprates is due to a very large transitional dipole moment between nearly
degenerate one- and two-photon states. Our findings suggest a strong
potential of these materials for high bit-rate all-optical switching and,
therefore, introduce a new means of achieving ultrafast optoelectronics with
strongly correlated electron systems. We believe that with innovative 
material processing there is considerable scope for
future enhancement of the figure of merit of these materilals for
optoelectronic applications.

This work was supported by a grant-in-aid for COE Research from the Ministry
of Education, Science, Sports, and Culture of Japan, and the U.S. NSF-ECS.

\begin{figure}[tbp]
\caption{(a) Crystal structure of ${\rm Sr_2CuO_3}$. (b) Absorption spectrum
of ${\rm Sr_2CuO_3}$, obtained by Kramers-Kronig transformation of
reflectivity. The solid triangles (i) -- (iv) indicate the pump photon
energies in (c). (c) Temporal evolution of $\Delta T / T$, measured at
various pump and probe photon energies at $\sim 10\;{\rm K}$. The temporal
evolution consists of two components: the hatched peak-like structure
corresponds to coherent interaction of two pulses, and the exponential tail
corresponds to incoherent charge-transfer excitation induced absorption. The
relaxation times $\protect\tau$ are the best fits to the exponential tails.}
\label{evo}
\end{figure}

\begin{figure}[tbp]
\caption{Spectra of photoinduced absorption efficiency $\protect\beta(
\protect\omega_{{\rm probe}};\protect\omega_{{\rm pump}})$, for pump
energies of (i) 0.7 eV, (ii) 1.1 eV, and (iii) 1.55 eV ($\sim 10\;{\rm K}$).
The solid lines are guides to the eye. (b) $\protect\beta$ vs. sum of the
pump and probe photon energy $\protect\omega_{{\rm pump}}+\protect\omega_{
{\rm probe}}$, for four different pump energies ($\sim 10\;{\rm K}$). The
dotted line is the linear absorption $\protect\alpha(\protect\omega)$, which
is plotted with the same energy scale as $\protect\omega_{{\rm pump}}+
\protect\omega_{{\rm probe}}$ for comparison.}
\label{probe_dep}
\end{figure}

\begin{figure}[tbp]
\caption{The calculated linear absorption and TPA spectra of our finite
periodic ring in arbitrary units (the TPA spectrum is plotted against $2
\protect\omega$), for $U_{Cu}$ = 10 eV, $U_O$ = 6 eV, $V$ = $\protect
\epsilon $= $t$ = 1 eV. The energies of the one- and two-photon states, and
that of the lowest singlet spin excitation m of the finite ring, relative to
the ground state G are shown as an inset. Notice the near degeneracy of the
one- and two-photon states. The transition dipole coupling $\protect\mu_{21}$
is an order of magnitude larger than the coupling $\protect\mu_{G1}$.}
\label{calc}
\end{figure}

\begin{figure}[tbp]
\caption{The transmission change induced by two temporally separated
equivalent pump pulses. The pump and probe photon energies are $0.88\;{\rm eV
}$ ($1.4\;{\rm \protect\mu m}$) and $1.03\;{\rm eV}$ ($1.2\;{\rm \protect\mu 
m}$), respectively. The relaxation time of the tail part is $1.2\;{\rm ps}$.
The transmission change with the second pump pulse is the same as that with
the first. The accumulation of the tail part is observed by comparing with
single-pulse response(not shown).}
\label{double}
\end{figure}


\begin{references}
\bibitem{comment on EMT}  Since the exciton in the strongly correlated
charge-transfer insulator has a very large binding energy, exciton
ionization can not occur at our pump intensities. This is in strong contrast
with conventional semiconductors, in which the exciton mote transition takes
place at a moderate excitations.
\end{references}
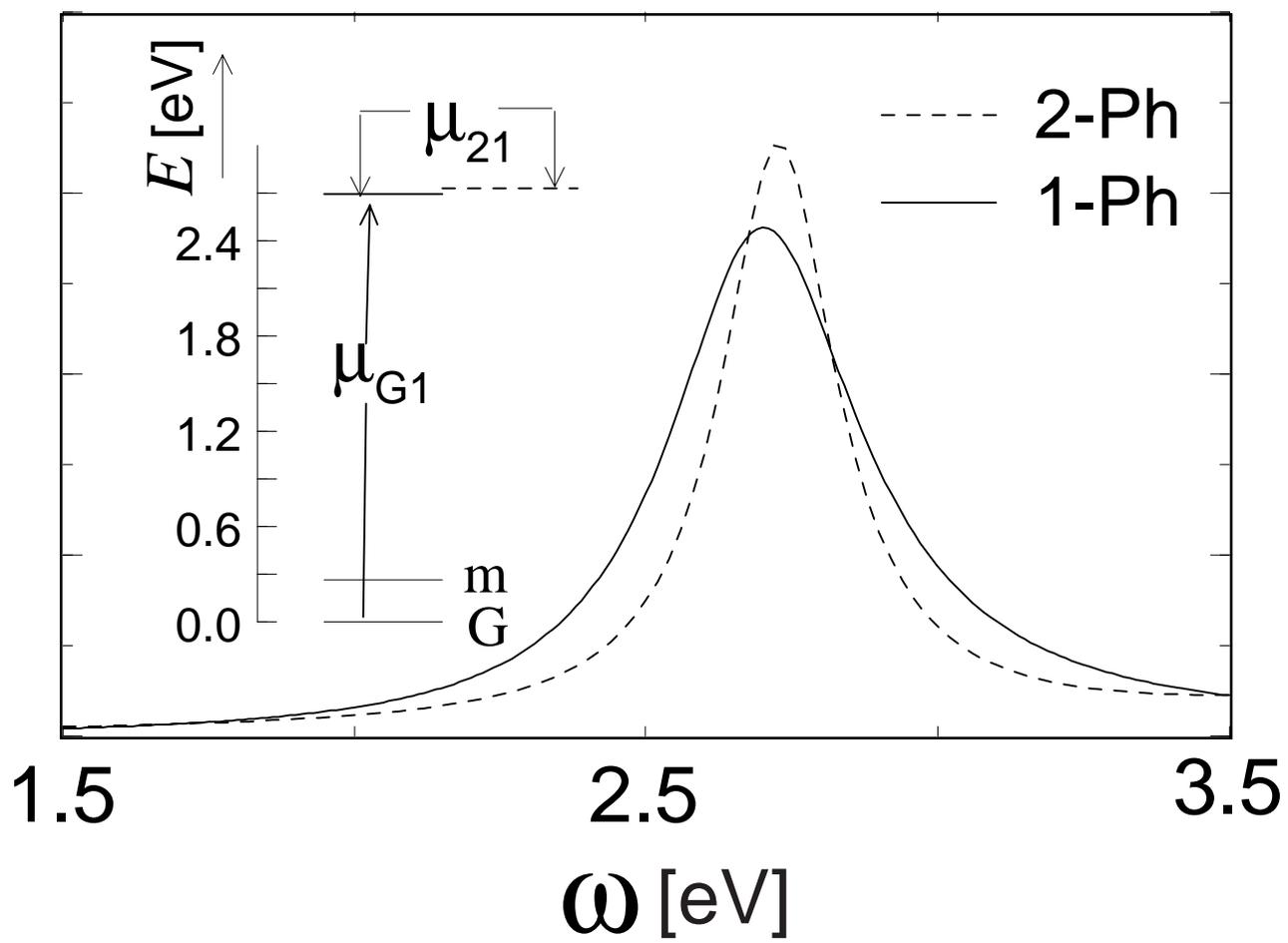
\end{document}